\documentclass{emulateapj}

\def\mo{M$_\odot$\,}
\def\sun{\hbox{$\odot$}}
\def\cm3{cm$^{-3}$}

\def\edot{$\dot{\rm E}$}
\def\msun{M$_{\odot}$}

\def\beq{\begin{equation}}
\def\eeq{\end{equation}}

\def\sles{\lower2pt\hbox{$\buildrel {\scriptstyle <}
   \over {\scriptstyle\sim}$}}
\def\sgreat{\lower2pt\hbox{$\buildrel {\scriptstyle >}
   \over {\scriptstyle\sim}$}}

\begin{document}

\title{The Proto-neutron Star Phase of the Collapsar Model and the Route to 
Long-soft Gamma-ray Bursts and Hypernovae} 
 
\author{L. Dessart\altaffilmark{1},
A. Burrows\altaffilmark{1},
E. Livne\altaffilmark{2},
and C.D. Ott\altaffilmark{1}
}
\altaffiltext{1}{Department of Astronomy and Steward Observatory,
                 The University of Arizona, Tucson, AZ \ 85721;
                 luc@as.arizona.edu,burrows@as.arizona.edu,cott@as.arizona.edu}
\altaffiltext{2}{Racah Institute of Physics, The Hebrew University,
Jerusalem, Israel; eli@frodo.fiz.huji.ac.il}

\begin{abstract}

  Recent stellar evolutionary calculations of low-metallicity massive fast-rotating 
main-sequence stars yield iron cores at collapse endowed with high angular momentum.
It is thought that high angular momentum and black hole formation are critical ingredients 
of the collapsar model of long-soft $\gamma$-ray bursts (GRBs). Here, we present 
2D multi-group, 
flux-limited-diffusion MHD simulations of the collapse, bounce, 
and immediate post-bounce phases of a 35-\msun\, collapsar-candidate model 
of Woosley \& Heger. We find that, provided the magneto-rotational instability (MRI)
operates in the differentially-rotating surface layers of the millisecond-period
neutron star, a magnetically-driven explosion ensues during the proto-neutron star
phase, in the form of a baryon-loaded non-relativistic jet, and that a black hole, 
central to the collapsar model, does not form.
Paradoxically, and although much uncertainty surrounds stellar mass loss, 
angular momentum transport, magnetic fields, and the MRI,
current models of chemically homogeneous evolution at low metallicity yield massive stars 
with iron cores that may have {\it too much} angular momentum to avoid 
a magnetically-driven, hypernova-like, explosion in the immediate post-bounce phase.
We surmise that fast rotation in the iron core may inhibit, rather than enable,
collapsar formation, which requires a large angular momentum not in the core but {\it above} it.
Variations in the angular momentum distribution of massive stars at core collapse 
might explain both the diversity of Type Ic supernovae/hypernovae and 
their possible association with a GRB.  
A corollary might be that, rather than the progenitor mass, the angular momentum distribution, 
through its effect on magnetic field amplification, distinguishes these outcomes.

\end{abstract}


\section{Introduction}

  There is mounting observational evidence for the association between long-soft $\gamma$-ray 
Bursts (GRBs) and broad-lined Type Ic supernovae (SNe; see Woosley \& Bloom 2006 for a review). 
Such hydrogen-deficient (and, perhaps, also helium-deficient) progenitors are 
compact and, if fast rotating in their 
core at collapse, fulfill critical
requirements for the formation of a collapsar (Woosley 1993).
The engine that converts energy from long-term accretion of disk material 
onto the black-hole (BH) may 
power a relativistic jet in the excavated polar regions.
The jet breaks out of the progenitor surface while equatorial accretion continues.
Depending on the BH mass and the angular momentum budget in the collapsing envelope,
this ``engine'' may operate for seconds, i.e. as long as typical long-soft GRBs.
Accompanying this beamed relativistic polar jet might be a disk wind, fueled by 
neutrinos or MHD processes, 
which would explode the Wolf-Rayet envelope. This explosion and the radioactive $^{56}$Ni 
material produced might lead to very energetic, broad-lined, Type Ic SN of 
the hypernova variety (Iwamoto et al. 1998; MacFadyen \& Woosley 1999;
Hjorth et al. 2003; Stanek et al. 2003).

   State-of-the-art radiation-hydrodynamics simulations including a sophisticated  
equation of state (EOS) and detailed neutrino transport (Buras et al. 2003; 
Burrows et al. 2006,2007a; Kitaura et al. 2006; Marek \& Janka 2007; Mezzacappa et al. 2007) 
suggest that while the neutrino mechanism of supernova explosions may work 
for the lower-mass massive progenitors, it may not for the more massive progenitors,
characterized by an ever higher post-bounce accretion rate onto the proto-neutron star (PNS). 
Burrows et al. (2006,2007a) have suggested that an acoustic mechanism will work for
all slowly rotating progenitors that do not explode by other means within the first 
second after bounce.
However, massive star cores endowed with a large angular momentum at the time of collapse 
should experience the magneto-rotational instability (MRI; Balbus \& Hawley 1991; 
Akiyama et al. 2003; Pessah et al. 2006; Shibata et al. 2006; Etienne et al. 2006), 
with the potential to exponentially amplify weak initial fields on a rotation timescale.
The saturation values of such fields are ultimately set by the free-energy of 
differential rotation available in the surface layers of the 
PNS (Ott et al. 2006), and can be large, i.e., on the order of 10$^{15}$\,G at 
a radius of a few tens of kilometers. 
The corresponding magnetic stresses at the neutron star surface lead systematically 
to powerful jet-like explosions $\sim$100\,ms after bounce 
(see, e.g., Ardeljan et al. 2005; Yamada \& Sawai 2004; Kotake et al. 2004; 
Sawai et al. 2005; Moiseenko et al. 2006; Obergaulinger et al. 2006;
Burrows et al. 2007b, hereafter B07; Dessart et al. 2007). 

   In this letter, we investigate, in the context of the collapsar model, the potential implications 
of this magnetic explosion mechanism. Our study focuses on the immediate post-bounce phase,
whose importance was emphasized by Wheeler et al. (2000,2002). This
is in contrast to previous work which explored only the phase subsequent to BH formation
(MacFadyen \& Woosley 1999; Aloy et al. 2001; Zhang et al. 2003; Proga 2005).
Indeed, two terms sometimes used in the collapsar context
are ``failed SN'' (MacFadyen \& Woosley 1999) and ``prompt BH formation'' 
(MacFadyen et al. 2001). Our analysis supports
the idea that the conditions for the collapsar model, as stated so far, are also suitable for a 
magnetically-driven explosion in the immediate post-core-bounce PNS phase, and that BH formation may 
be so delayed for a range of putative progenitor models that it does not in fact 
occur\footnote{In the present context, BH formation is never prompt, since it takes a finite time,
on the order of seconds, for the PNS to accumulate the critical mass at which it
experiences the gravitational instability. This is in contrast with super-massive 
stars, such as the progenitors of pair-instability SNe, which may form an apparent horizon 
during collapse and thus ``directly'' transition to a BH (Liu et al. 2007).}.
In \S2, we present radiation MHD simulations with the code VULCAN/2D 
(Livne et al. 2004,2007) of a collapsar-candidate model that support this thesis. 
In \S3, we discuss the implications of our results for stellar evolutionary models
that might lead to collapsars and/or hypernovae.

\begin{deluxetable}{lcccccccc}
\tablewidth{0pt}
\tabletypesize{\scriptsize}
\tablecaption{Properties of our two MHD-VULCAN/2D simulations
of the 35OC collapsar model of WH06.\label{tab_model}}
\tablehead{
\colhead{}&
\colhead{$t_{\rm end}$}&
\colhead{$t_0$}&
\colhead{M$_{10}$}&
\colhead{P$_{10}$}&
\colhead{E$_{\rm expl}$}&
\colhead{\edot$_{\rm gas}$}&
\colhead{\edot$_{\rm \vec{E} \times \vec{B}}$}&
\colhead{$v_{\rm max}$}\\
\colhead{}&
\colhead{ms}&
\colhead{ms}&
\colhead{\mo}&
\colhead{ms}&
\colhead{B}&
\multicolumn{2}{c}{B\,s$^{-1}$}&
\colhead{km\,s$^{-1}$}
}
\startdata
M0 & 369 & ...  & 2.1 &  4 & 0.03 & 0.5 & 0.25 & 43,000 \\
M1 & 666 & 349  & 1.7 & 12 & 3.31 & 9.4 & 3.0  & 58,000 \\
\enddata
\tablecomments{
$t_{\rm end}$ gives the time at the end of each simulation, while $t_0$ is the time 
when the rate of polar mass ejection first overcomes equatorial mass accretion.
All quoted quantities in the table correspond to the final time in each simulation,
while times are given with respect to core bounce. 
M$_{10}$ (P$_{10}$) corresponds to the total baryonic mass (average rotation
period) inside the 10$^{10}$\,g\,cm$^{-3}$ 
isodensity contour. \edot$_{\rm gas}$ (\edot$_{\rm \vec{E} \times \vec{B}}$)
is the Bernoulli (Poynting) power in the ejecta, obtained by integrating 
the corresponding flux over a shell with a radius of 500\,km. 
[See text for additional information.]
}
\end{deluxetable}

\section{Model and results}

   We present results from two-dimensional, rotating, multi-group, flux-limited diffusion
magneto-hydrodynamics simulations, using VULCAN/2D (Livne et al. 2004,2007; see also
appendices in Dessart et al. 2006 and Burrows et al. 2007a), 
of the 35OC progenitor model and collapsar candidate of 
Woosley \& Heger (2006; hereafter WH06). The numerical procedure we follow is identical 
to that of B07 in every respect, except the choice of progenitor.
WH06's model is evolved from a 35\,\msun\, zero-age main sequence star endowed
with a total angular momentum of 1.4$\times$10$^{53}$\,erg\,$\cdot$\,s, a metallicity 
of 1\% the solar value, and a reduction
by a factor of 10 in the prescribed mass loss rates during the Wolf-Rayet phase.
Our simulations extend out to a maximum radius of 5000\,km (which contains $\sim$3\,\msun) 
and cover a 90$^{\circ}$ quadrant, bounded by the rotation axis and the equator.
We adopt WH06's initial rotational, density, temperature, and electron-fraction profiles
for that progenitor.
For the magnetic field distribution, we start with magnitudes and morphology that
are consistent with the 35OC model of WH06.
We use a field uniform within 3000\,km, and dipolar beyond. 
In our reference model, M0, we employ initial poloidal and toroidal field magnitudes
of 2$\times$10$^{10}$\,G and 8$\times$10$^{11}$\,G, respectively, 
in close quantitative agreement with the 35OC model of WH06. 
However, we also study a model, M1, with an initial poloidal field that is five times stronger. 
This leads to a magnetic field energy at $\sim$100\,ms after core bounce 
that is closer to the value expected at the PNS surface, were we to adequately resolve 
the MRI (B07).
In Table~1, we give important quantities characterizing the two simulations performed.
Note that if magnetic fields are ignored in the pre-collapse evolutionary calculations 
of WH06, core angular velocities reach 5-22\,rad\,s$^{-1}$, much larger than the
1.98\,rad\,s$^{-1}$ achieved in the 35OC model. A bounce at sub-nuclear 
densities may ensue and lead to BH formation (Akiyama \& Wheeler 2005). 
Ignoring magnetic torques in the models most prone to magnetic-field generation
during the pre-collapse phase seems inconsistent. Thus, we focus on the 
more slowly rotating progenitors, evolved with magnetic fields, which inevitably
bounce at nuclear densities.

\begin{figure*}
\includegraphics[width=.3\textwidth]{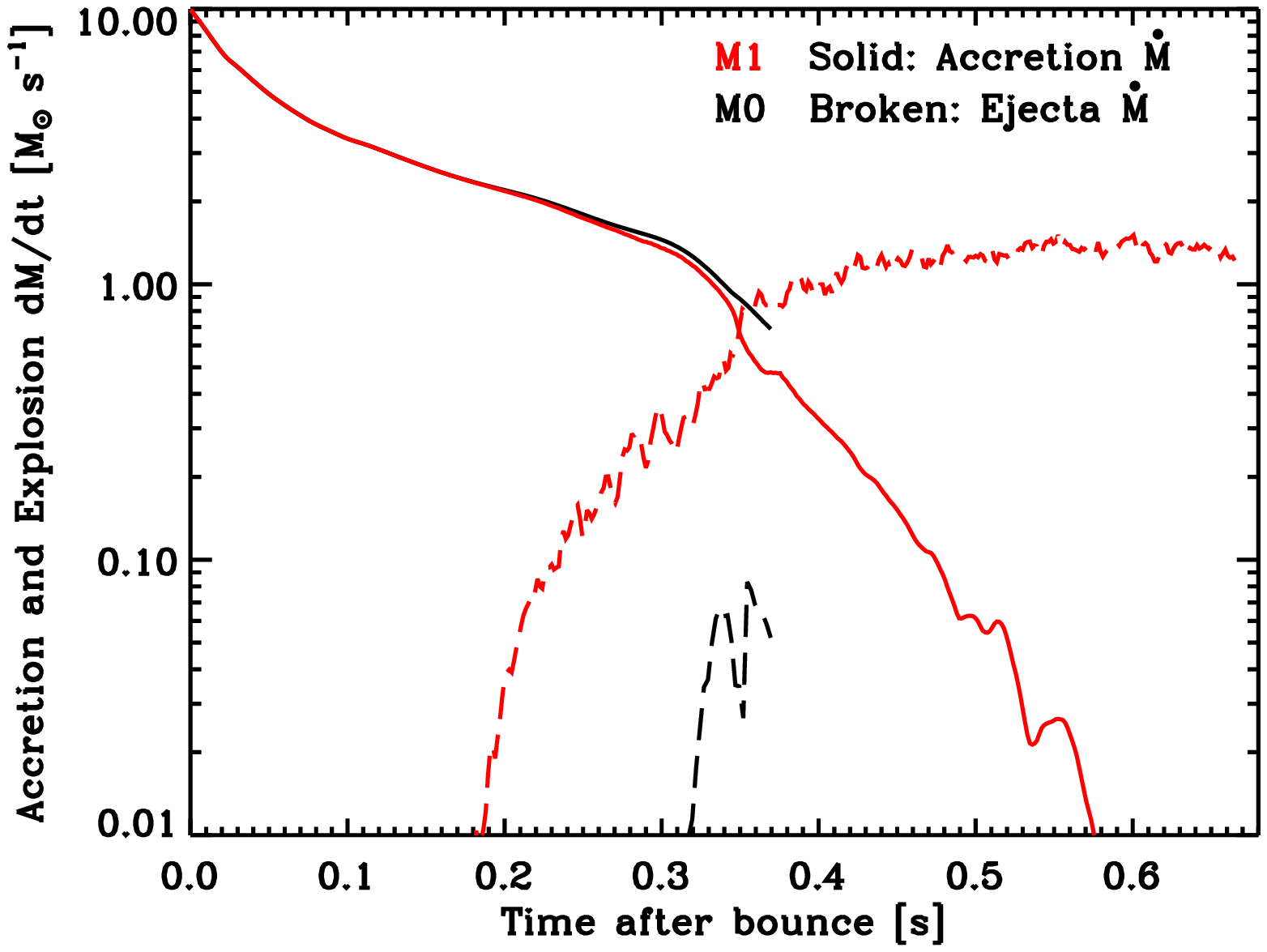}
\includegraphics[width=.3\textwidth]{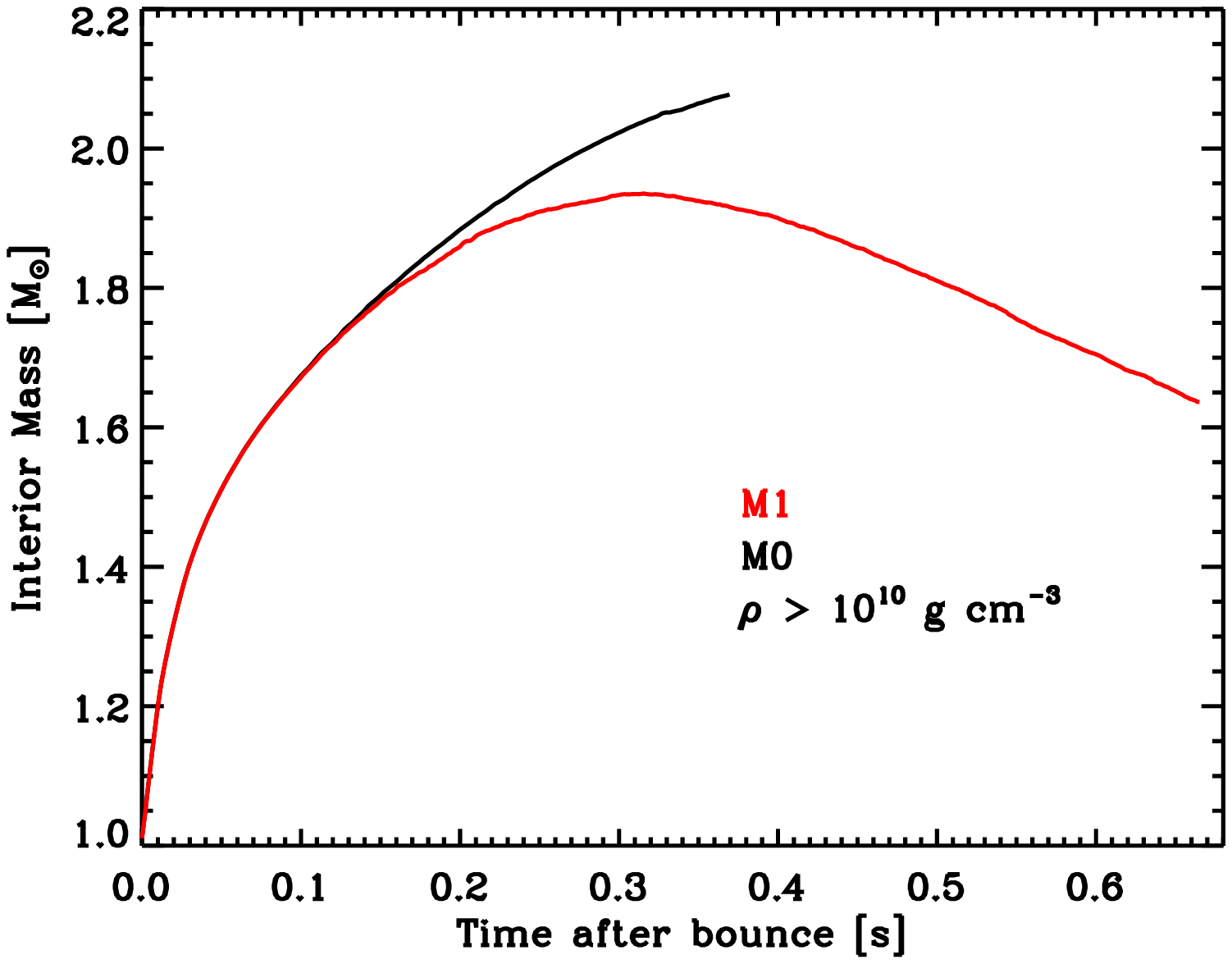}
\includegraphics[width=.3\textwidth]{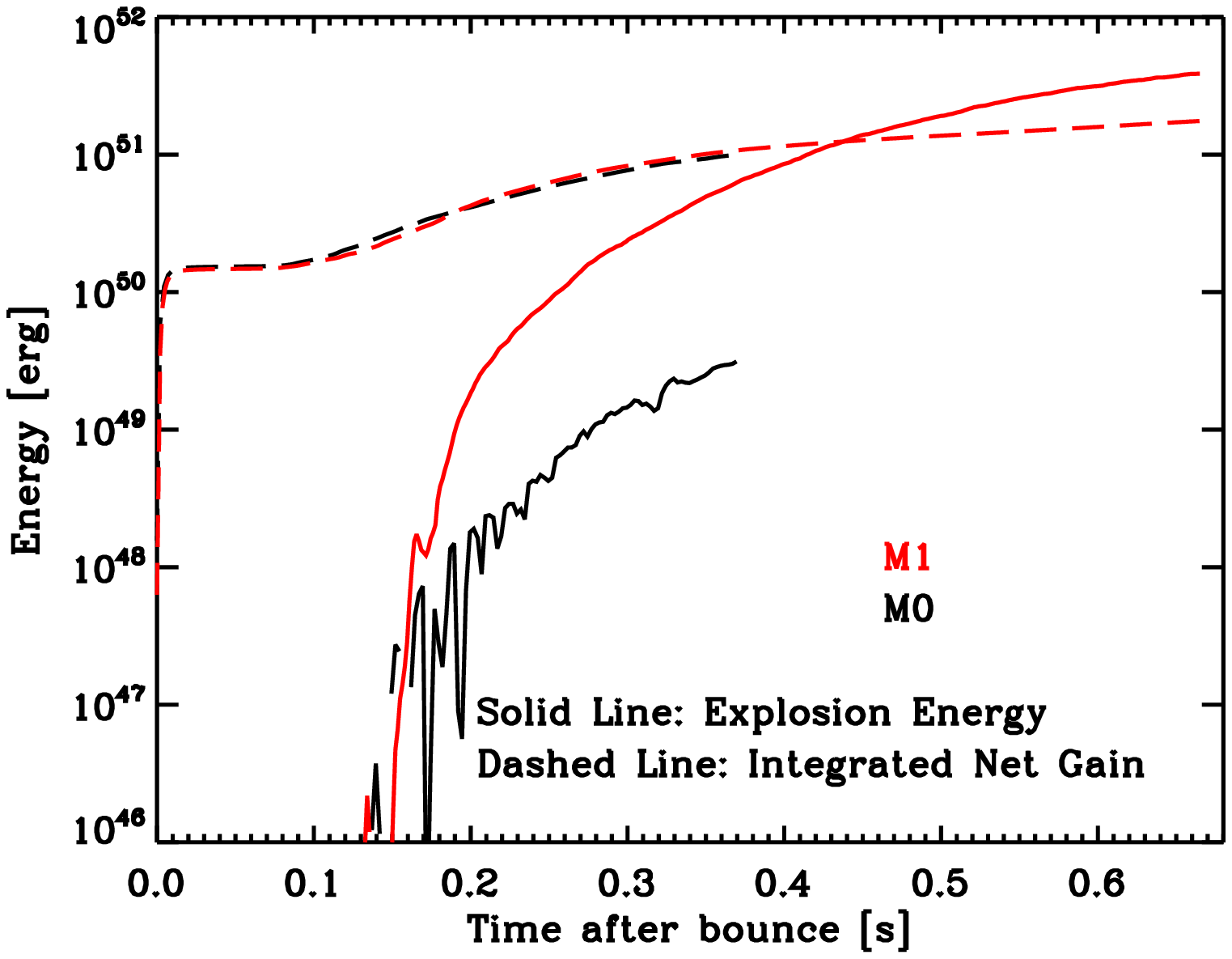}
\caption{
{\it Left:} Time evolution of the instantaneous integrated  mass flux
accreting ({\it solid line}) and outflowing ({\it  dashed lines}) through a shell at a radius
of 500\,km, for models M0 (black) and M1 (red). 
{\it Middle:} Same as left, but for the total mass interior to the iso-density
contour corresponding to 10$^{10}$\,g\,cm$^{-3}$.
{\it Right:} Same as left, but for the explosion energy ({\it solid line}) and 
the net integrated neutrino gain ({\it  dashed lines})
outside the high-density regions bounded by the 10$^{10}$\,g\,cm$^{-3}$ contour.
[See text for discussion.]
}
\label{fig_prop}
\end{figure*}

\begin{figure}
\includegraphics[width=8cm]{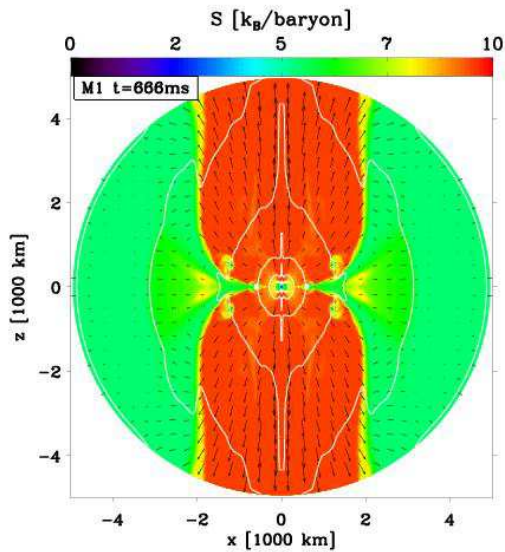}
\caption{Colormap of the entropy at 666\,ms after bounce for model M1,
overplotted with white iso-density contours (every decade downward from
10$^{10}$\,g\,cm$^{-3}$) and velocity vectors (length saturated to 15\% of the width
of the display and corresponding to a velocity of 30,000\,km\,s$^{-1}$.)
}
\label{fig_still}
\end{figure}

   After an initial collapse phase that lasts $\sim$245\,ms, the central density reaches 
$\sim$3$\times$10$^{14}$\,g\,cm$^{-3}$, the EOS stiffens, a shock is born, and
propagates outward, but is debilitated by the 
photo-dissociation of the infalling outer iron-core and the burst of electron neutrinos.
The shock stalls at $\sim$150\,km, and within a few tens of ms after bounce, it becomes
increasingly aspherical. The net gain from neutrino emission and absorption processes, the entropy,
and the material accretion rates get progressively larger at larger latitudes as the degree of 
oblateness of the fast-rotating PNS increases. Subsequent to the
amplification due to compression by a factor of $\sim$2500 in both magnetic field components, 
the toroidal magnetic field increases after bounce due to the winding of the poloidal field 
component. At the same time, accretion of the outer magnetized core continues and 
enhances the total magnetic field energy. By 150\,ms (300\,ms) after bounce, the 
magnetic pressure at the surface of the PNS along the pole in the M1 (M0) model
is comparable to the gas pressure, and a bipolar, magnetically-driven, baryon-loaded, and 
non-relativistic jet is initiated, reversing accretion into ejection along the polar direction. 
As shown in the left panel of Fig.~\ref{fig_prop}, the initial
jet mass-loss rate is only $\sim$0.01\,M$_{\odot}$\,s$^{-1}$, but in the M1 model and
by $\sim$350\,ms after bounce it {\it exceeds}
the accretion rate. At this time, the accumulated baryonic PNS mass 
is only 1.93\,\msun\, (middle panel of Fig.~\ref{fig_prop}), 
and, thus, well below the 2.17\,\msun\, baryonic mass transition to BH 
formation that we derive with our Shen EOS (Shen et al. 1998). 
Note that for a solid-body rotator this limit may increase by up to 10-20\%
(Cook et al. 1994), but for a differentially-rotating neutron star this limit may be considerably
larger (Baumgarte et al. 2000). Hence, rotation, and in particular differential rotation, 
enhances the potential for explosion during a PNS phase. 
By 650\,ms after bounce, the explosion energy in 
the M1 model reaches $\sim$3.3\,B (10$^{51}$\,erg $\equiv$ 1\,Bethe [1\,B]) 
(see right panel of Fig.~\ref{fig_prop}), although due to the continued 
accretion along near-equatorial latitudes a quasi-steady-state is reached with 
an explosion power sustained at $\sim$10\,B\,s$^{-1}$. 
As shown in Fig.~\ref{fig_still}, the jet resembles 
a magnetic tower (Lynden-Bell 2003; Uzdenski \& MacFadyen 2001), but is confined primarily 
by the ram pressure of the infalling dense envelope (B07).
As time progresses, its base broadens and, given the quasi-steady jet conditions, the 
mass ejection rate grows and the accretion is limited to progressively smaller latitudes. 
Both cause and effect,
the increase in ejecta volume enhances the neutrino contribution to the explosion energy,
although it remains a subdominant part of the total by the end of the simulation.
Extraction of core rotational energy by magnetic torques 
is also in evidence in the M1 model from the increase in the average 
PNS rotation period\footnote{We define the average rotation-period 
as the period of the rigidly-rotating PNS that has the same total angular momentum 
and structure inside the 10$^{10}$\,g\,cm$^{-3}$ isodensity contour.\label{footnote_period}} 
from 8 to 12\,ms between 200\,ms and 600\,ms after bounce, while over half this interval
the average period decreases from 5 to 4\,ms in the weakly exploding M0 model. 
The decrease in the free energy of rotation in the M1 model is on the order of 
3\,B, and is comparable to the magnitude of the 
explosion energy. This supports the idea that core rotation energy fuels the 
magnetically-driven ejecta.
Hence, the M1 model, modified slightly to yield fields at saturation that
agree roughly with what would obtain in the presence of the MRI, boasts a clear and 
powerful explosion.
In this model, and once the explosion is well established, the PNS loses mass 
at a steady rate, and has a mass of only $\sim$1.7\,\msun\, at the end of the simulation.
The broadening of the base of the jet suggests that the explosion will not choke
(nor induce any significant fallback), 
and encompassing a larger solid angle, will instead lead to explosion in all directions. 
Hence, such an object is unlikely ever to transition to a BH and to lead to a collapsar. 

  By contrast, in the M0 model, the explosion emerges later, when the neutron star baryonic mass has 
already accumulated $\sim$2.1\,\msun, and thence may be susceptible to collapse to a BH. 
The free-energy of core rotation has been partially tapped, but the potential subsequent 
powering of a GRB may not be compromised. This model, by mimicking more slowly rotating cores
or an inefficient MRI, offers a limiting case for the formation or non-formation of a BH, 
and a possible collapsar.


\section{Discussion}

  The potential for exponential growth on a rotational timescale 
of initial seed magnetic fields by the MRI (Shibata et al. 2006; Etienne et al. 2006),
fueled by the free energy of core rotation,
makes the initial angular momentum budget of the progenitor star the key parameter 
in determining the outcome during the immediate post-bounce phase (B07).
A magnetically-driven, baryon-loaded, and non-relativistic, 
explosion is obtained for WH06's 35OC collapsar candidate model,
evolved at low metallicity from a 35\,\msun\, fast-rotating main-sequence star. 
The explosion occurs $\sim$200\,ms after bounce and reaches $\sim$3\,B
$\sim$400\,ms later. 
After an initial accretion phase, the steadily decreasing PNS mass reaches
only $\sim$1.7\,\msun\, at the end of the simulation,
and, thus, the quasi-steady explosion we observe suggests that BH formation is unlikely to occur.
Moreover, baryon contamination prevents the ejecta from becoming relativistic. 
Note that the production of a GRB in the collapsar context is contingent on the gravitational 
collapse of the PNS to a BH.

The recent stellar evolutionary calculations of Yoon \& Langer (2005),
WH06, and Meynet \& Maeder (2007) of fast-rotating main-sequence objects at low metallicity
systematically predict such fast-rotating cores at collapse. 
Starting from similar conditions
for a 35-40\,\msun\, star, but using different mass-loss ``recipes,'' 
they obtain very similar rotational profiles in the inner core. 
Allowing for anisotropic mass loss (Meynet \& Maeder 2007),
a model of C. Georgy (2007, priv. comm.) suggests an even larger (by a factor of two) 
specific angular momentum in the inner 3\,\msun\, at the end of silicon core burning. 
Despite the agreement between these different evolutionary computations,
the magnetically-driven explosion and the ``failed'' 
BH formation described here are conditional on the uncertain treatment of mass loss, 
angular momentum transport, and magnetic processes (Spruit 2002) during the pre-collapse evolution.

At very low metallicities, radiatively-driven winds of massive stars are  
inhibited by the lack of metals (Kudritzki 2002; Vink et al. 2001; Vink \& de Koter 2005), 
whose optically-thick lines intercept radiation momentum (Castor et al. 1975). 
Recent revisions downward of mass-loss rates due to 
clumping (Owocki et al. 1988; Bouret et al. 2005; Fullerton et al. 2006)
suggest, however, the potential importance of episodic outbursts, 
akin to the 1843 giant eruption of Eta Carina (Smith \& Owocki 2006). 
The metallicity dependence of such phenomena is
entirely unknown, mostly  because the fundamental cause of the outburst remains a mystery.
While line driving seems excluded, continuum driving of a porous medium at super-Eddington
luminosities has been proposed by Owocki et al. (2004) as an alternative. 
Finally, mass loss in fast-rotating, and sometimes critically-rotating (Townsend et al. 2004), 
envelopes is complicated by the effects associated with centrifugal support, surface oblateness,
and gravity-darkening (Cranmer \& Owocki 1995; Owocki et al. 1996), so that the mass-loss 
``recipes'' used in stellar evolutionary 
models are not always substantiated by observational and theoretical 
evidence.
At present, and in light of our simulations, it appears that chemically-homogeneous evolution
of fast-rotating main-sequence massive stars at low metallicity systematically 
yields iron cores at collapse that may have {\it too much} angular momentum, a property 
that prevents the formation of a collapsar. Uncertainties in the modeling of
the pre-collapse evolution may result, however, in slower-rotating iron cores\footnote{Note that the
rotational energy $E_{\rm rot}$ is a stiff function of angular velocity $w$, i.e., 
$E_{\rm rot} \propto w^2$.} and, thus, might inhibit an early magnetically-driven explosion 
in favor of black hole, and perhaps collapsar, formation.

We conclude that variations in the angular momentum distribution of pre-collapse
massive stars may lead to different post-bounce scenarios. Non- or slowly-rotating progenitors
may explode with weak/moderate energy ($\sles$1\,B) through a neutrino 
or an acoustic mechanism $\sles$1\,s after bounce, or may collapse to a BH.
Objects with large angular momentum in the envelope, but little in the core,
may proceed through the PNS phase, transition to a BH and form a collapsar
with a GRB signature. Owing to the modest magnetic-field amplification above the
PNS, a weak precursor polar jet may be launched, soon overtaken by a baryon-free, 
collimated relativistic jet. At the same time, the progenitor envelope  is exploded by a disk wind, 
resulting in a hypernova-like SN with a large luminosity (large $^{56}$Ni mass).
Finally, and this is what we conclude here, objects with large angular momentum in the core 
may not transition to a BH.
Instead, and fueled by core-rotation energy, a magnetically-driven
baryon-loaded non-relativistic jet is obtained without any GRB signature. 
The explosion has the potential of reaching energies of a few B to 10\,B, 
and for viewers along the poles of looking like a Type Ic hypernova-like SN with broad lines. 
For a viewer at lower latitudes, the delayed and less energetic explosion
nearer the equator may look more like a standard Type Ic SN 
(H\"{o}flich et al. 1999).
This volume-restricted jet-like explosion is dimmer, as the
amount of processed $^{56}$Ni may be significantly less than the $\sim$0.5\,\msun\,
obtained in the collapsar context (MacFadyen \& Woosley 1999).
Hence, magnetic processes during the post-bounce phase of fast-rotating iron cores offer 
a potential alternative to collapsar formation and long-soft GRBs
by producing non-relativistic non-Poynting-flux-dominated baryon-loaded 
hypernova-like explosions without any GRB signature.
Importantly, while our study narrows the range over which the collapsar model may exist, it
also offers additional routes to explain the existence of GRB/SN-hypernova events 
like SN 1998bw (Woosley et al. 1999), and hypernova events like SN 2002ap 
without a GRB signature (Mazzali et al. 2002).

 More generally, magnetic effects should naturally arise in the context of gravitational collapse
and fast rotation. The resulting angular momentum of newly-formed BHs and magnetars, for example, 
would be reduced, perhaps considerably, by any prior magnetically-driven explosion, and, thus, 
may decrease the power of subsequent mass ejections from compact objects 
(see, e.g., Thompson et al 2004).

\acknowledgments

We thank Cyril Georgy, Norbert Langer, and Sung-Chul Yoon for
providing their stellar evolutionary models,
for a comparison with the 35OC model of WH06 used in this work. 
We also thank Stan Woosley for fruitful discussions and insight. 
We acknowledge support for this work
from the Scientific Discovery through Advanced Computing
(SciDAC) program of the DOE, under grant numbers DE-FC02-01ER41184 
and DE-FC02-06ER41452, and from the NSF under grant number AST-0504947.
E.L. thanks the Israel Science Foundation
for support under grant \# 805/04, and C.D.O. thanks 
the Joint Institute for Nuclear Astrophysics (JINA) for support under
NSF grant PHY0216783. This research used resources of the National 
Energy Research Scientific Computing Center, which is supported by the
Office of Science of the U.S. Department of Energy under Contract No. DE-AC03-76SF00098.


\end{document}